\documentclass[aps,prd,showpacs,nofootinbib]{revtex4}
\usepackage{dcolumn}
\usepackage{amsmath,amssymb,setspace,graphicx,subfigure,epsfig}
\usepackage{amsfonts}
\usepackage{latexsym}
\usepackage{epsfig}
\usepackage{color}
\usepackage{nicefrac}

\newcommand{\dis}[1]{\begin{equation}\begin{split}#1\end{split}\end{equation}}

\newcommand{\be}{\begin{equation}}
\newcommand{\ee}{\end{equation}}
\newcommand{\eq}[1]{Eq.~(\ref{#1})}

\newcommand{\Mp}{M_P}

\newcommand{\p}{\partial}
\newcommand{\vp}{\varphi}
\newcommand{\vps}{\varphi_*}
\newcommand{\chis}{\chi_*}

\newcommand{\ep}{\epsilon_\vp}
\newcommand{\epe}{\epsilon_\vp^e}

\newcommand{\ec}{\epsilon_\chi}
\newcommand{\ece}{\epsilon_\chi^e}

\newcommand{\fnl}{f_{\rm NL}}

\def\tr{\tilde{r}}

\newcommand{\calP}{{\cal P}}

\def\sis{\sin^2\theta^*}
\def\etap{\eta_{\varphi\varphi}}
\def\etac{\eta_{\chi\chi}}

\def\bea{\begin{eqnarray}}
\def\eea{\end{eqnarray}}

\def\Np{N_{\varphi}}
\def\Nc{N_{\chi}}
\def\Npp{N_{\varphi\varphi}}
\def\Npc{N_{\varphi\chi}}
\def\Ncc{N_{\chi\chi}}
\def\Npcc{N_{\varphi\chi\chi}}
\def\Nccc{N_{\chi\chi\chi}}
\def\Ncccc{N_{\chi\chi\chi\chi}}
\def\fNLt{f_{NL}^{\rm{tree}}}
\def\fNLl{f_{NL}^{1\;\;\rm{loop}}}
\def\Prat{\frac{P_{\zeta}^{1loop}}{P_{\zeta}^{tree}}}
\def\Pz{\mathcal{P}_{\zeta}}

\begin{document}
\title{Large non-Gaussianity from two-component hybrid inflation}
\author{Christian T. Byrnes}
\email{C.Byrnes@thphys.uni-heidelberg.de}
\affiliation{Institut f\"ur Theoretische Physik, Universit\"at Heidelberg, Philosophenweg 16, 69120
Heidelberg, Germany}
\author{Ki-Young Choi}
\email{kiyoung.choi@uam.es}
\affiliation{Departamento de F\'{\i}sica Te\'{o}rica C-XI,
        Universidad Aut\'{o}noma de Madrid, Cantoblanco,
        28049 Madrid, Spain}
\affiliation{Instituto de F\'{\i}sica Te\'{o}rica UAM/CSIC,
        Universidad Aut\'{o}noma de Madrid, Cantoblanco,
        28049 Madrid, Spain}
\author{Lisa M.H. Hall}
\email{lisa.hall@sheffield.ac.uk}
\affiliation{Department of Applied Mathematics, University of Sheffield, Sheffield, S3 7RH, UK}


\hfill HD-THEP-08-31, FTUAM 08/22, IFT-UAM/CSIC-08-83

\date{\today}

\begin{abstract}
We study the generation of non-Gaussianity in models of hybrid inflation with two
inflaton fields, (2--brid inflation). We analyse the region in the parameter and the initial
condition space where a large
non-Gaussianity may be generated during slow-roll inflation which is generally characterised by a
large $\fnl$, $\tau_{NL}$ and a small $g_{NL}$. 
For certain parameter values we can satisfy $\tau_{NL}\gg\fnl^2$. 
The bispectrum  is of the local type but may have a significant scale dependence. 
We show that the loop corrections to the power spectrum and
bispectrum are suppressed during inflation, if one assume that the fields follow a classical
background trajectory. We also include the effect of the waterfall field, which can lead to a
significant change in the observables after the waterfall field is destabilised, depending on
the couplings between the waterfall and inflaton fields.
\end{abstract}

\maketitle


\section{Introduction}

The increasingly accurate observations of the cosmic microwave 
background~\cite{Komatsu:2008hk} motivate the study of
inflation beyond linear order in perturbation theory. Since there are many models of inflation
which give similar predictions for the spectral index of the scalar perturbations and the
tensor-to-scalar ratio~\cite{Liddle:2000cg,non-gaussian1} it is important to consider higher order
observables. Non-Gaussianity has emerged
as a powerful discriminant between different models of inflation. Any detection of primordial
non-Gaussianity would rule out the simplest models of single field inflation.

Many ways to generate a large non-Gaussianity have been proposed in the
literature~\cite{Alabidi:2006hg,Bartolo:2001cw,Enqvist:2004ey,cline,Dutta:2008if,Misra:2008tx}. If the
inflaton has a non-canonical kinetic term then the inflaton field perturbations may already be
non-Gaussian
at Hubble exit during inflation~\cite{DBI}. Otherwise the perturbations are Gaussian at Hubble exit
and any
non-Gaussianity must then be generated on super Hubble scales 
\cite{Maldacena:2002vr,Seery:2005gb}. Popular methods to generate a large
non-Gaussianity after inflation include having a feature in the
inflaton potential~\cite{Chen:2006xjb}, the curvaton scenario~\cite{curvaton}, modulated
reheating/preheating~\cite{modulatedreheating,Suyama:2007bg},
an inhomogeneous end 
of inflation~\cite{Bernardeau:2002jy,endofinflation,Alabidi:endinf,Sasaki:2008uc,Naruko:2008sq}
or by loop domination of the bispectrum or trispectrum~\cite{Cogollo:2008bi,Rodriguez:2008hy}.
Work has also been done on calculating the non-Gaussianity during inflation with a separable
potential \cite{Vernizzi:2006ve,Choi:2007su,Battefeld:2006sz,Seery:2006js} and with more
general potentials~\cite{Rigopoulos:2005us,Yokoyama:2007uu}.
Recently the authors of this paper have shown that it is 
possible to generate a large non-Gaussianity during multiple field inflation while keeping all of
the slow-roll parameters much less than unity~\cite{Byrnes:2008wi}.

In this paper we study this possibility of generating a large non-Gaussianity during slow-roll
inflation in detail for the specific model of hybrid inflation. We consider carefully the parameter
constraints and initial conditions required to do this. Unlike in the previous work~\cite{Byrnes:2008wi} we here also
consider the effect of the waterfall field required to end inflation. We can therefore also
compare and contrast our work to a recent paper on generating a large non-Gaussianity at the
end of inflation~\cite{Naruko:2008sq}. Depending on the values of the couplings between the two inflaton fields and
the waterfall field observable quantities may change when the waterfall field is destabilised. This
change in the primordial curvature perturbation is possible because there are isocurvature
perturbations present at the end of inflation.

We also calculate further observables for this model, such as the trispectrum and the scale
dependence of
the bispectrum. In most cases the trispectrum is large through a large $\tau_{NL}$ whenever the
bispectrum is
large but the relation between them also depends on the initial conditions. 

Recently it has also been shown that for two-field hybrid inflation it is possible to generate a
large loop correction to the bispectrum and trispectrum~\cite{Cogollo:2008bi,Rodriguez:2008hy}. For
a special trajectory it was shown that the bispectrum
can be observably large even when the tree level value of the bispectrum is slow-roll suppressed.
Here we show that this is not possible for a classical trajectory, 
since we require that the classical motion of the field
dominates over its quantum fluctuations in order that a slow-roll calculation is valid.

The plan of our paper is as follows:
In Section~\ref{sechybrid} we review the generation of a large bispectrum
during hybrid inflation and calculate the trispectrum.
In Section~\ref{sec:waterfall} we include the waterfall field  
and consider its effect at the end of inflation for different couplings
between the waterfall and inflaton fields and further evolution after 
inflation. In Section~\ref{sec:scale}
we calculate the scale dependence of the non-Gaussianity parameter.
We conclude in Section~\ref{sec:conclusion}.
In the Appendix, we show that the loop correction is subdominant to the tree level 
term for the power spectrum, bispectrum and trispectrum.

\section{Non-Gaussianity during hybrid inflation}
\label{sechybrid}


We consider a model of two field hybrid inflation, whose potential is given by
\bea\label{Whybrid}
W(\vp,\chi)= 
 W_0\left(1+\frac12\etap\frac{\vp^2}{\Mp^2}+\frac12\etac\frac{\chi^2}{\Mp^2}\right), \eea
which is vacuum dominated, i.e.~which satisfies $\left|\etap\vp^2\right|\ll\Mp^2$ and
$\left|\etac\chi^2\right|\ll\Mp^2$. 
We assume that inflation ends abruptly by
a waterfall field which is heavy during inflation and hence doesn't affect the dynamics during
inflation. In this section we calculate observables during slow-roll inflation. We will consider
the full potential including the waterfall field in Sec.~\ref{sec:waterfall}. We will see that this
can lead to a change in observables on the surface where the waterfall field is destabilised.

In the vacuum dominated regime the slow-roll solutions are 
\dis{\label{hybridsoln}
\vp= \vp_*e^{-\etap N}, \qquad \chi= \chi_*e^{-\etac N},
}
where `*' denotes the value at the horizon exit. Throughout this section whenever we write a
quantity without making it explicit at which time it should be evaluated, we mean the equation to
be valid at any time $N$ $e$--foldings after Hubble exit and while slow roll is valid. Generally we
will be interested in quantities at the end of inflation, in which case we take $N=60$.

The slow-roll parameters are
\dis{\label{hybridsr}
\epsilon_\vp=\frac12\etap^2 \frac{\vp^2}{\Mp^2}
,\qquad
\epsilon_\chi=\frac12\etac^2 \frac{\chi^2}{\Mp^2}
, \qquad \epsilon= \epsilon_\vp+ \epsilon_\chi.
}
We note that the dominant slow-roll parameters $\etap$ and $\etac$ are
constants during inflation in the vacuum dominated regime and that they are much
larger than the slow-roll parameters $\epsilon_\vp$ and 
$\epsilon_\chi$ throughout inflation.

Using the $\delta N$ formalism~\cite{starob85,ss1,Sasaki:1998ug,lms,lr} (for details see the appendix)
we can calculate the power spectrum
$\calP_\zeta$, spectral index $n_\zeta$, tensor-to scalar ratio
$r$, where $\calP_T$ is the power spectrum of tensor perturbations, and the non-linearity parameter
$\fnl$ in this
model~\cite{Bellio-Wands96,Alabidi:2006hg,Vernizzi:2006ve,Choi:2007su,Byrnes:2008wi}, at leading order,
\dis{
\calP_\zeta=\frac{W_*}{24\pi^2\Mp^4\epsilon^2}\left(\ep e^{-2\etap N}+\ec e^{-2\etac N}
\right),\label{Pzeta_hybrid}
}

\bea
n_\zeta -1 &=& -2\epsilon^* + 2\frac{(\etap-2\epsilon e^{2\etap N} )\ep e^{-2\etap N}
+(\etac-2\epsilon e^{2\etac N})\ec e^{-2\etac N} }{\ep e^{-2\etap
N}+\ec e^{-2\etac N}}\nonumber  \\
 &\simeq& 2\frac{\etap\ep e^{-2\etap N} +\etac\ec e^{-2\etac N} }{\ep e^{-2\etap
N}+\ec e^{-2\etac N}}\label{tilt}, \eea

\dis{
r\equiv \frac{\calP_T}{\calP_\zeta} = \frac{16\epsilon^2}{\ep e^{-2\etap N}+\ec e^{-2\etac N}},
}
\dis{
\fnl = \frac56 \frac{-\epsilon\left[\etap\epsilon_\vp +\etac\epsilon_\chi e^{4(\etap-\etac)N}
\right] + \frac{2}{\epsilon}\epsilon_\vp\epsilon_\chi(\etap\epsilon_\chi + \etac \epsilon_\vp
)\left[1-  e^{2(\etap-\etac)N}\right]^2}{\left[\epsilon_\vp+\epsilon_\chi e^{2(\etap-\etac)N} 
\right]^2}.\label{fnl_hybrid}
}  
According to~\cite{Byrnes:2008wi}, large non-Gaussianity can be realised
in either of two regions
\dis{
\cos^2\theta \equiv\frac{\dot{\vp}^2}{\dot{\vp}^2+\dot{\chi}^2}\simeq
\frac{\epsilon_\vp}{\epsilon_\vp+\epsilon_\chi } \ll 1,\qquad\textrm{or}\qquad 
\sin^2\theta \equiv \frac{\dot{\chi}^2}{\dot{\vp}^2+\dot{\chi}^2}\simeq
\frac{\epsilon_\chi}{\epsilon_\vp+\epsilon_\chi } \ll 1. 
}
Since both regions are symmetrical~\cite{Byrnes:2008wi} (before specifying the values of $\etap$
and $\etac$), in the rest of this paper we will focus on the second
region (Region B or D in Ref.~\cite{Byrnes:2008wi}). In this region
where $\epsilon_\vp\gg\epsilon_\chi$,
$|\fnl|>1$ is fulfilled by the condition, 
\bea \sis\lesssim \sin^4\theta \left(\sqrt{\frac{5|\etac|}{6\sin^2\theta}}-1\right)\, ,
 \label{fnlB}
\eea
in other words,
\dis{
|\etac|^{-1}e^{-4(\etap-\etac)N}  \lesssim \sin^2\theta  \simeq
\frac{\epsilon_\chi}{\epsilon_\vp}\lesssim |\etac|.
}
This condition implies three conditions on the parameter $\theta$:
\bea 
\sis<\frac13\left(\frac{5}{6}\right)^2\left(\frac{3}{4}\right)^4\left| \eta_{\chi\chi} \right|^2,
\quad \sin^2\theta<\frac{5}{6}\left| \eta_{\chi\chi} \right|, \qquad
\frac{\sin^2\theta}{\sis}>\frac{24}{5}\frac{1}{|\etac|}.\label{threeconditions}
\eea

Note that in this region $\sin\theta\simeq\etac\chi/(\etap\vp)$, from Eq.~(\ref{hybridsoln}) we
require $N(\etap-\etac)>1$ so
that $\sin^2\theta$ grows significantly during inflation.

\subsection{Simplified formula for the observables when $\fnl$ is large}

We can substantially simplify all of the above formula in the case where $\fnl$ is large. We define
the quantity 
\dis{\label{trdefinition}
\tr\equiv \left(\frac{\p N}{\p \chi_*}\right)^2/ \left(\frac{\p N}{\p
\vp_*}\right)^2=\frac{\ec}{\ep} e^{2(\etap-\etac)N}.
}
In the region we are considering where $\fnl$ is large,
this is approximately given by the initial and final angles of the background trajectory with
different exponents
\bea \tr\simeq \frac{\sin^4\theta}{\sis}. \eea
We note that $\tr$ can be either larger or smaller than one.

In the case of large non-Gaussianity it follows that
\dis{
\calP_\zeta\simeq\frac{W_*}{24\pi^2\Mp^4\epsilon_*}\left(1+\frac{\ec}{\ep}e^{2(\etap-\etac)N}
\right)= \frac{8}{r}\left(\frac{H_*}{2\pi}\right)^2,\label{Pzeta_our}
}
\dis{
n_\zeta -1 \simeq 2\frac{\etap +\tr \etac}{1+\tr},
}
\dis{
r\simeq \frac{16\epsilon^*}{1+\tr},
}
\dis{\label{fNL_our}
\fnl\simeq
 \frac56 \frac{\sin^6\theta^e}{(\sis+\sin^4\theta^e)^{2}}\etac 
=\frac56\frac{\tr }{(1+\tr)^2}
\etac e^{2(\etap-\etac)N}.
}

The first condition  in Eq.~(\ref{threeconditions})  implies that
\bea \frac{\chis}{\vps}\ll1.  \eea
We therefore require a very small value of $\chis$ in order to have a 
large non-Gaussianity. We will see in sec~\ref{loop} that there is a limit
to how small we can make $\chis$ while $\chi$ still follows
a classical slow-roll trajectory. In practise for relatively large values of
 $\etap-\etac$ this places a bound on how large we can make $\fnl$.

The sign of $\fnl$ is determined by the sign of $\etac$. 
The amplitude of $\fnl$ depends exponentially on the difference of the 
slow-roll parameters, $\etap-\etac$, which we
require to be positive to be in the branch of large non-Gaussianity 
where $\sin^2\theta\ll1$.
However the spectral index depends on the weighted sum of the slow-roll 
parameters, so it is
possible to have a large non-Gaussianity and a scale invariant spectrum. However it is not
possible to have a large and positive $\fnl$ and a red spectrum of perturbations. We will see
in Sec.~\ref{subsecB} that by including the effect of the waterfall field this conclusion may
change, depending on the values of the coupling constants between the two inflaton fields and the
waterfall field.

Using the $\delta N$ formalism we can also calculate the non-linearity parameters which
parameterise the trispectrum. There are two shape independent
parameters, which may be
observationally distinguishable and are given by~\cite{Byrnes:2006vq} 
\bea\label{NLformula} 
\fnl=\frac56\frac{N_{AB}N_AN_B}{(N_CN_C)^2}, \qquad \tau_{NL}=\frac{N_{AB}N_{BC}
N_AN_C}{\left(N_DN_D\right)^3},\qquad
 g_{NL}=\frac{25}{54}\frac{N_{ABC}N_AN_BN_C}{\left(N_DN_D\right)^3},\eea
where summation over the fields is implied over the repeated indices $A,B,C,D$. 
For comparison we also give the formula for $\fnl$
\cite{lr}. In the
regime where $|g_{NL}|>1$ it is given by
\bea  g_{NL}=\frac{10}{3}\frac{\tr\left(\etap-2\etac\right)-\etac}{1+\tr}\fnl. \eea
Details of the calculation for both $g_{NL}$ and $\tau_{NL}$ are in the appendix, in particular see
Eqs.~(\ref{firstd}),~(\ref{secondd}) and~(\ref{N3rd}). We see that $g_{NL}$ is subdominant to
$\fnl$ and
hence won't provide a competitive observational signature. 
The current observational bound on
the local type of the bispectrum from WMAP data is $-9<\fnl<111$ at the $2\sigma$ level
\cite{Komatsu:2008hk} and currently there is no observational constraint on the trispectrum. If
there is no detection it is expected that with Planck data the bounds will be reduced to about
$|\fnl|\lesssim10$ and $\tau_{NL}\lesssim560$ \cite{Kogo:2006kh}.

However in the regime where $\tau_{NL}$ is large, which is similar to the region
where $\fnl$ is large we find
\bea\label{NLtrispectrum}
\tau_{NL}=\frac{\tr}{(1+\tr)^3}\etac^2e^{4N(\etap-\etac)} = \frac{1+\tr}{\tr}
\left(\frac{6}{5}\fnl\right)^2. 
\eea
From Eq.~(\ref{fNL_our}) we see that for given model parameters $\etap$ and $\etac$ the value
of
$\fnl$ is maximised for $\tr=1$, i.e. when we choose the initial field values to
satisfy
\bea \frac{\chis}{\vps}=\frac{\etap}{\etac}e^{-2N(\etap-\etac)}. \eea
In this case the non-linearity parameters are given by
\bea  \fnl=\frac{5}{24}\etac e^{2N(\etap-\etac)},\qquad
\tau_{NL}=2\left(\frac65 \fnl\right)^2.  \eea
However $\tau_{NL}$ is maximised at a slightly different point, for $\tr=1/2$.

It follows from (\ref{NLtrispectrum}) that $\tau_{NL}> (6\fnl/5)^2$, so
$\tau_{NL}$ may be large and provide an extra observable parameter for this model. This
inequality between $\tau_{NL}$ and $\fnl$ is true in general \cite{Suyama:2007bg}, and equality is
reached whenever a single field direction during inflation generates the 
primordial curvature perturbation. However it is usually assumed that $\tau_{NL}\sim\fnl^2$
since both arise from second derivatives in the $\delta N$ formalism. In fact for our model it is
possible to have a small $\fnl$ (and hence also a small $g_{NL}$) but a large and potentially
observable $\tau_{NL}$. For this we require that $\tr\ll1$, in practice if we make it too small it
may no longer be possible to satisfy the classical constraint (\ref{classicalconstraint}) discussed
in Sec.~\ref{loop}. In the final example in Table~\ref{table_hybrid} we give an explicit
example of parameter values which give rise to an $\fnl$ which is probably too small to be detected
with Planck but with a very large trispectrum through $\tau_{NL}>10^3$ that should be detectable at
a high significance. 
For another example with $f_{NL},g_{NL}\lesssim O(1)$ but $\tau_{NL}\gg1$, see
~\cite{Ichikawa:2008ne}.
In contrast it has been shown in several papers
\cite{Enqvist:2005pg,Byrnes:2006vq,Huang:2008bg,Huang:2008zj,Enqvist:2008gk} that in the curvaton
scenario where the curvaton has a non-quadratic potential it is possible to realise $|g_{NL}|\gg1$
while $\tau_{NL}=(6\fnl/5)^2$ is small with some tuning of parameters. Recently it has also been
shown that in a single field model with a non-canonical kinetic term it is possible for certain
tuned parameters to achieve a much larger trispectrum than bispectrum \cite{Engel:2008fu}. We note
that the non-Gaussianity in their model has a very different shape dependence to the $k$ independent
non-linearity parameters we are considering in this paper.

In Table~\ref{table_hybrid}, we give some explicit examples of values of $\etap,\,\etac,\,\vps$ and
$\chis$ which
lead to a large non-Gaussianity. Using Eq.~(\ref{tilt}) we also calculate the
spectral index. 
The contours of $\fnl$ of potential for a specific choice of $\etap$ and
$\etac$ is given
in Fig.5 in our previous paper~\cite{Byrnes:2008wi}. 
We correct tensor-to-scalar ratio in Table 1 of~\cite{Byrnes:2008wi}.
\footnote{The value of the tensor-to-scalar ratio in Table 1 of 
\cite{Byrnes:2008wi} is incorrect. The correct
value for the first row is 0.005, for the second row is 0.026 (as given in Table \ref{table_hybrid})
and for the third row $r=0.002$. This does not change the discussion or conclusions of
\cite{Byrnes:2008wi}.}
The first example in the Table~\ref{table_hybrid} shows that it is possible to have
$|\fnl|\simeq100$ and a scale invariant spectrum. 
We also see that it is possible to generate a large non-Gaussianity during slow roll
with $\etap$ and $\etac$ both positive or both negative, or when one is positive and the other negative corresponding to a saddle point.
%
\begin{table}
\begin{tabular}{|cc|cc|cc|c|c|c|c|c|c|}
\hline
$\etap$ & $\etac$ & $\vps$ & $\chis$ &$\vp_e$&$\chi_e$& $\tr$ & $\fnl$ & $\tau_{NL}$ & $g_{NL}$ & $n_{\zeta}-1$ &   
r \\
\hline0.04 & -0.04   & 1 & 6.8$\times10^{-5}$&0.091&$7.50\times10^{-5}$  & 1 & -123  & 4.4$\times10^{4}$ & -33 & 0 & 0.006 \\ 
0.04 & -0.04 & 1 & $1.5\times10^{-4}$&0.091&$1.65\times10^{-3}$ & 5 & -68 & $8\times10^{3}$ & -24 & -0.05 & 0.002 \\
0.08 & 0.01    & 1 & 0.0018     &0.008& $9.88\times10^{-4}$ & 1 & 9.27  & 247   & 0.77  & 0.09 & 0.026  \\ 
0.02 & -0.04   & 1 & 0.00037    &0.301&$4.08\times10^{-3}$  & 1 & -11.1 & 357   &-2.6 &-0.02 & 0.002 \\
-0.01 & -0.09 &  1 & $3\times10^{-6}$&1.822& $6.64\times10^{-4}$& 0.16 & -132 & $1.8\times10^{5}$ & -44 & -0.04 & 0.0007  \\
0.06 & -0.01 & 1 & $4.3\times10^{-4}$&0.027& $7.84\times10^{-4}$& 0.1 & -3 & 148& -0.2& 0.11 & 0.026    \\
0.01 & -0.06& 1& $7.5\times10^{-6}$&0.549& $2.75\times10^{-4}$& 0.04 & -8 & 2.5$\times 10^{3}$ & -2 & 0.01 & 0.0008  \\
\hline
\end{tabular}
\caption{Table showing some initial conditions for the hybrid inflation model that lead to large
levels of non-Gaussianity. The end point of fields, $\tr$, the bispectrum and trispectrum non-linearity parameters, spectral index and tensor-to-scalar ratio are shown. 
They are evaluated when the number of e-foldings from the end of inflation 
is $N_k=60$.
}
\label{table_hybrid}
\end{table}

\subsection{Can the loop correction dominate?}
\label{loop}

Cogollo {\it et al.~}have calculated the effect of the loop correction to the
primordial power spectrum and bispectrum~\cite{Cogollo:2008bi} and even more recently Rodriguez and
Valenzuela-Toledo have calculated the loop correction to the trispectrum~\cite{Rodriguez:2008hy}.
In both cases this was for the special case of a straight background
trajectory where one of the fields is zero throughout inflation. This loop correction arises from
taking into account the contribution to the power spectrum and bispectrum arising from terms in the
$\delta N$ expansion which are non-leading in the expansion of the field perturbation $\delta\vp_*$,
see e.g.~\cite{Byrnes:2007tm}. However they can still be significant if the coefficient to the term
given by the second derivative of $N$ with respect to the subdominant field is extremely large
and the leading term for the same field is small or zero,
e.g.~\cite{Boubekeur:2005fj}. These ``higher order" terms are usually neglected, but
\cite{Cogollo:2008bi} has shown the first explicit example
of an inflation model where they cannot be neglected. They consider a 2-field hybrid model with the
same potential as Eq.~(\ref{Whybrid}) in the special case of an unstable straight trajectory along
one of the axes, with $\etap$ and $\etac$ both negative. In this case, they find (for certain
initial values) that one of the loop correction terms is dominant over the tree
level term and can generate an observable $\fnl$. 

However if the value becomes too small then the motion of the $\chi$ field will become
dominated by quantum
fluctuations rather than the classical drift down the potential,
 $3H\dot{\chi}\simeq-W_{\chi}=-W_0\etac\chi/M_P^2$, which we have assumed.
In order that we can neglect the effect of the quantum fluctuations, the condition we require on the
background trajectory is \cite{Creminelli:2008es}
\bea\label{classicalconstraint} |\dot{\chi}|\pi/H^2>\sqrt{3/2}. \eea
We require the condition above to be satisfied throughout inflation.
Since the fields in our model either increase or decrease monotonically it is sufficient to check
that the constraint is satisfied both at Hubble exit and at the end of inflation. 
In the appendix
we show that satisfying this condition requires that the loop corrections are suppressed. We have
checked that for the examples in Table~\ref{table_hybrid} this condition is satisfied.
We stress that this suppression of the loop corrections follows from the requirement that the
background trajectory of the inflation fields is dominated by the classical motion. However it would
be interesting to investigate the large loop corrections found in
\cite{Cogollo:2008bi,Rodriguez:2008hy} using a calculation which includes the quantum corrections to
the $\chi$ field, which is set to zero throughout inflation in
\cite{Cogollo:2008bi,Rodriguez:2008hy}.

Because we have shown that the loop correction is always suppressed compared to the tree
level terms we are justified in using the formula for the tree-level non-Gaussianity parameters as
given by Eq.~(\ref{NLformula}).

\section{Non-Gaussianity after hybrid inflation}\label{sec:waterfall}

In this section we include the effects of the waterfall field $\rho$ which is required to end hybrid
inflation. Inflation ends when the waterfall field is destabilised, i.e.~when its effective mass
becomes negative. During inflation the waterfall field is heavy and it is trapped with a vacuum
expectation value of
zero, so we can neglect it during inflation.
The end of inflation occurs when the effective mass of the waterfall field is zero, which occurs on
a hypersurface defined in general by \cite{Sasaki:2008uc,Naruko:2008sq},
\dis{
\sigma^2=G(\vp,\chi)\equiv g_1^2 \vp^2+g_2^2\chi^2,\label{endinflation}
}
which is realised by the potential $W(\vp,\chi)$, defined by Eq.~(\ref{Whybrid}), where $W_0$ is given
by
\dis{
W_0=\frac12G(\vp,\chi)\rho^2 + \frac{\lambda}{4}\left(\rho^2-\frac{\sigma^2}{\lambda} \right)^2.
}
Here $g_1$ ($g_2$) is the coupling between the $\vp$ ($\chi$) field and the waterfall field.
In general the hypersurface defined by this end condition is not a surface of uniform energy
density. Because the $\delta N$ formalism requires one to integrate up to a surface of uniform
energy density we need to add a small correction term to the amount of expansion up to the surface
where the waterfall field is destabilised, which we will consider in sec.~\ref{subsecC}. 

We note here that the hybrid potential, which we have written in the form of a sum potential can
also be written as a product potential in the limit of vacuum domination. We show this so that we
can use the formula from \cite{Naruko:2008sq}, we consider a potential of the form 
\bea\label{Whybridexp}
W(\vp,\chi)=
W_0\exp\left(\frac12\etap\frac{\vp^2}{\Mp^2}\right)\exp\left(\frac12\etac\frac{\chi^2}{\Mp^2}
\right).
\eea
We note that the slow-roll parameters are not exactly the same from these two ways of writing
the potential, but that we are only interested in calculating results at leading order in slow
roll, and in this limit the two ways of writing the hybrid potential are equivalent.

This is an example of a model with an inhomogeneous end of inflation, 
i.e.~where inflation ends at
slightly different times in different places. This has been studied as
 a method for converting the
inflationary isocurvature perturbation into the primordial curvature perturbation
\cite{endofinflation}.
It has also been shown for the hybrid potential we are considering 
that this can be used to generate a large amount of non-Gaussianity, 
for certain parameters values and fine tuning of the parameters
~\cite{Naruko:2008sq,Alabidi:endinf}. However these papers 
concern the large non-Gaussianity generated
 at the end of inflation rather than during slow-roll inflation, by having a very
large ratio of couplings $g_1/g_2\ll1$. Here we consider the case where $g_1$ and $g_2$ have the
same order of magnitude with $g_1^2/g_2^2=\etap/\etac$ in sec.~\ref{subsecA}
 and  with $g_1^2= g_2^2$ in sec.~\ref{subsecB}.

\subsection{$g_1^2/g_2^2=\etap/\etac$}
\label{subsecA}
\begin{figure}[!t]
  \begin{center}
 \begin{tabular}{c c}
   \includegraphics[width=0.4\textwidth]{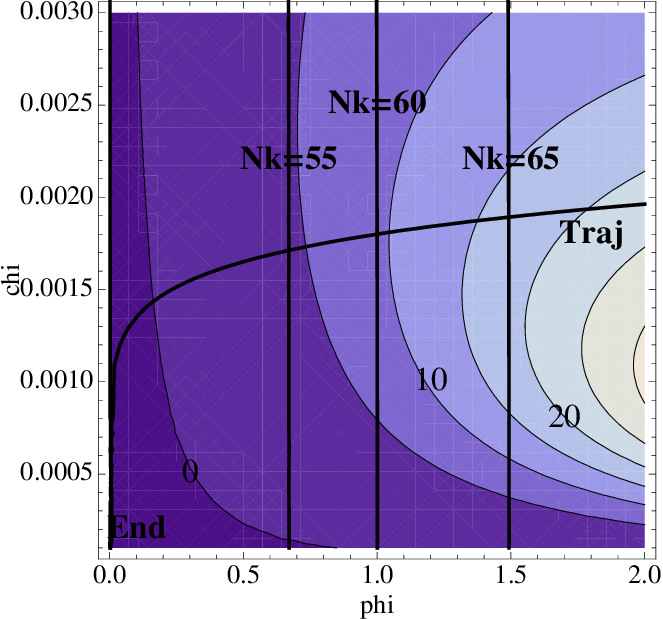} 
&
\includegraphics[width=0.4\textwidth]{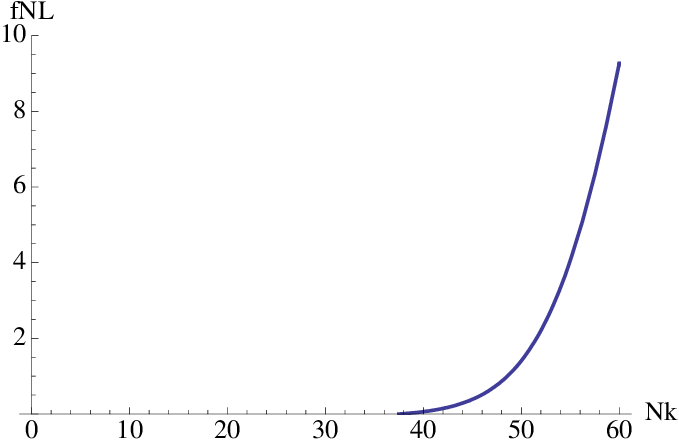} 
   \end{tabular}
  \end{center}
 \caption{Left: The contour plot of $\fnl$ which have the end condition same as uniform energy
density hypersurface $g_1^2/g_2^2=\etap/\etac$ in \ref{subsecA} in the plane of phi and chi, which
denote the values of the
fields when the given scale leaves the
horizon. The values of $\fnl$, 0, 10 and 20, are shown on the corresponding contour.
For example with the given end point, the trajectory is shown as a line denoted by
``Traj''. The cross point with $N_k=60$ line is the value of phi and chi when the
scale corresponding to $N_k=60$ leaves horizon. In this scale the value of $\fnl$ is around 9.
Right: Directly from the left figure, we can read the scale dependence of $\fnl$ on the given
trajectory. For the given trajectory the spectrum of $\fnl$ is shown in the figure. For this
trajectory we used the third example in  
Table~\ref{table_hybrid}, i.e. $\etap=0.08$, $\etac=0.01$, and the end point is
fixed so that the field values $\vp_*=1$ and $\chi_*=0.0018$ lead to the number
of e-foldings $N_k=60$.}\label{hybridA}
\end{figure}

In this case we have chosen the coupling constants (which can satisfy $g_1^2<0$ and/or $g_2^2<0$)
such that the surface where the waterfall field is destabilised corresponds to a surface of uniform
energy density. In this case the value of all observable quantities such as the power
spectrum and non-Gaussianity are the same as those we calculated previously which were valid at the
final hypersurface of uniform energy density during inflation. 

With these values of the coupling constants, large non-Gaussianity can only be generated
with the special conditions we have outlined earlier in this paper and it is given by \eq{Pzeta_our} -- \eq{fNL_our}. We have checked that our result for $\fnl$, Eq.~(\ref{fnl_hybrid}) in this case is consistent
with the formula for $\fnl$ in~\cite{Naruko:2008sq}. This provides a check on the algebra.
We emphasise that this large non-Gaussianity is not from the end of inflation
but from the evolution during inflation. To compare between our paper and \cite{Naruko:2008sq} we
note that in their notation $m_1^2=\etap$, $m_2^2=\etac$, $\epe=-m_1^3m_2\vp_e\chi_eW/(2Z)$ and
$\ece=m_1m_3^3\vp_e\chi_e Y/(2X)$ where the parameters $W,X,Y$ and $Z$ are defined in
\cite{Naruko:2008sq}. 

\begin{figure}[!t]
  \begin{center}
  \begin{tabular}{c c}
   \includegraphics[width=0.4\textwidth]{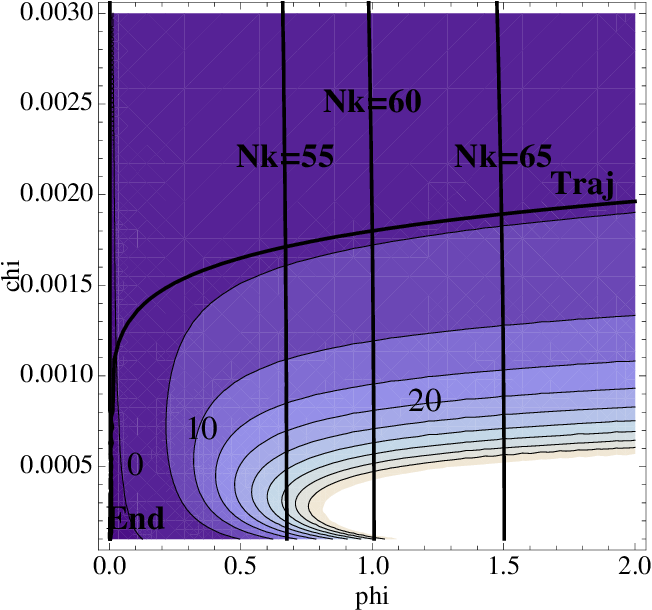} 
&
   \includegraphics[width=0.4\textwidth]{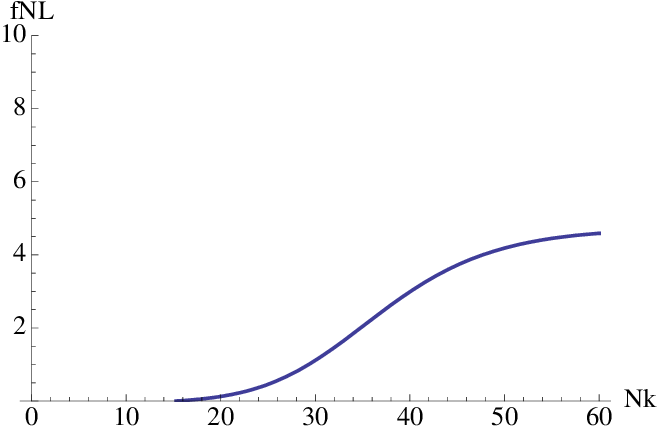} 
   \end{tabular}
  \end{center}
 \caption{Same as Fig.~\ref{hybridA} but using the end condition $g_1^2=g_2^2$, see
\ref{subsecB}.}\label{hybridB}
\end{figure}

In the left hand plot of figure~\ref{hybridA}, we show the contour plot 
of $\fnl$ in the $(\vp,\chi)$
plane of the field values when the given scale leaves the horizon.
$\fnl$ increase as the number of e-folding from the end of inflation increases, which is shown in
the right figure.
This can be easily understood from Eq.~(\ref{fNL_our}). 
For a small e-folding number, $N_k$, (corresponding to small scales) $\tilde{r}$ is also 
small and $\fnl$ is small. 
As $N_k$ increases, $\tilde{r}$ increases as well as $\fnl$.
However for $\tilde{r}\gtrsim 1$, which corresponds to 
$N_k\gtrsim [\log(\ep/\ec)]/ 2(\etap-\etac)$ for a given end point,
$\fnl$ approaches its the maximum value,
which corresponds to
\dis{
\fnl^{max}=\frac{5}{6}\frac{\epsilon_\vp}{\epsilon_\chi}\etac.
\label{fnlmaxA}
}
In the example given in Fig.~\ref{hybridA}, this happens 
around $N_k\gtrsim 120$, thus the maximum is not shown in the right-hand plot 
of Fig.~\ref{hybridA} in the given range of $N_k$.
The value of $N_k$ which gives the maximum value can be modified
by the condition of end of inflation and can be approached $N_k\sim60$ 
which is shown in the next section and Fig.~\ref{hybridB}.

\subsection{$g_1^2= g_2^2$}
\label{subsecB}

In this case, the end of inflation given by the
condition in Eq.~(\ref{endinflation}) does not occur on a uniform 
energy density hypersurface~\cite{Naruko:2008sq}. In this subsection we will show how the
non-Gaussianity is modified by the condition at the end of inflation with this specific
example. In general we expect there to be some modification to non-Gaussianity from the end of
inflation, except in the special case we considered in~\ref{subsecA}.

In this case the power spectrum and the non-linearity parameters are~\cite{Naruko:2008sq}
\dis{
\calP_\zeta\simeq \frac{W_0}{24\pi^2\Mp^4}\frac{(\ep/\etap^2)e^{-2\etap N}+(\ec/\etac^2)e^{-2\etac
N}}{(\ep/\etap+\ec/\etac)^2},
}
\dis{
\fnl= \frac56\frac{-\left(\ep/\etap^2+\ec/\etac^2e^{4(\etap-\etac)N} \right)
\left(\ep/\etap + \ec/\etac
\right)+ 2\frac{\ep\ec}{\etap^2\etac^2}\frac{\ep/\etap^2+\ec/\etac^2}{\ep/\etap+\ec/\etac}(\etac-
\etap e^{2(\etap-\etac)N})^2 }{\left[\ep/\etap^2+\ec/\etac^2e^{2(\etap-\etac)N}\right]^2}.
}

In the large $\fnl$ limit,
\dis{
\calP_\zeta\simeq\frac{W_*}{24\pi^2\Mp^4\epsilon_*}\left(1+\frac{\etap^2}{\etac^2}\frac{\ec}{\ep}e^{
2(\etap-\etac)N} \right)= \frac{8}{r}\left(\frac{H_*}{2\pi}\right)^2,\label{Pzeta_M}
}
\dis{
n_\zeta -1 \simeq 2\frac{\etap +\frac{\etap^2}{\etac^2}\tr \etac}{1+\frac{\etap^2}{\etac^2}\tr},
}
\dis{
r\simeq 16\epsilon^* \left(1+\frac{\etap^2}{\etac^2}\tr\right)^{-1},
}
\dis{\label{fNL_M}
\fnl\simeq
\frac56\frac{\frac{\etap^2}{\etac^2}\tr
}{\left(1+\frac{\etap^2}{\etac^2}\tr\right)^2} \etap
e^{2(\etap-\etac)N}.
}
We can see that the observables at the end of inflation are changed by the additional ratio
of $\etap^2/\etac^2$ before $\tr$. 
Furthermore, $\fnl$ has an additional factor of $\etap/\etac$.

In Table~\ref{table_hybridB} we show the values of $\fnl$, $n_{\zeta}-1$ and $r$ for the same
parameter values as we used in Table~\ref{table_hybrid}. The first two examples show that if
$\etap=-\etac$ then the observables are unchanged except that the sign of $\fnl$ is switched. The
second example in the table shows that in this case it is possible to have a red spectral index and
a positive value of $\fnl$. For many values of the initial parameters when $|\etap/\etac|\neq1$ the
magnitude of $\fnl$ decreases compared to Table~\ref{table_hybrid}. We see that this is the case
for most examples given in the table. This is most strikingly illustrated in the fifth example in
the two tables. We can see by comparing (\ref{fNL_our}) and (\ref{fNL_M}) that in the case where
$\tr\ll1$ and $|\etap/\etac|\ll1$ that the magnitude of $\fnl$ will decrease by about
$|\etap/\etac|^3$. However for most values of $\tr$ and the ratio $\etap/\etac$ the change is much
smaller. In some cases $\fnl$ increases, by the largest amount when $\tr<1$ and $|\etap/\etac|>1$.
An example of this where $|\fnl|$ grows by more than an order of magnitude is shown in the sixth
column of Tables~\ref{table_hybrid} and \ref{table_hybridB}.

Due to the additional factor before $\tilde{r}$, $\fnl$ is increased at small
scales compared to the case in sec.~\ref{subsecA} and 
decreased at large scales due to the additional factor. Thus the maximum 
$\fnl$ is smaller than that of  sec.~\ref{subsecA} by $\etap/\etac$.
At small scales (where $\tr \ll 1$), ${\fnl}_A= (\etap/\etac){\fnl}_B$.
At large scales (where $\tr \gg 1$),  ${\fnl}_A= (\etac/\etap){\fnl}_B$.

\begin{table}
\begin{tabular}{|cc|cc|c|c|c|c|}
\hline
$\etap$ & $\etac$ & $\vps$ & $\chis$ & $\tr$ & $\fnl$  & $n_{\zeta}-1$ & r \\
\hline

0.04 & -0.04   & 1 & 6.8$\times10^{-5}$  & 1 & 123   & 0 & 0.006 \\ 
0.04 & -0.04 & 1 & $1.5\times10^{-4}$ & 5 & 68  & -0.05 & 0.002 \\
0.08 & 0.01 & 1 & 0.0018 & 1 & 4.59 &0.02 & 0.0008  \\ 
0.02 & -0.04 & 1 & 0.00037 & 1 & 3.5 &0.02 & 0.026 \\
-0.01 & -0.09 &  1 & $3\times10^{-6}$ & 0.16 & -0.2 & -0.02 & 0.0008  \\
0.06 & -0.01 & 1 & $4.3\times10^{-4}$ & 0.1 & 38 &0.01 & 0.006    \\

\hline
\end{tabular}
\caption{Same as Table~\ref{table_hybrid} but with different end condition, $g_1^2=g_2^2$ as 
used in sec.~\ref{subsecB}. }
\label{table_hybridB}
\end{table}

\subsection{The end of hybrid inflation}

\label{subsecC}
As we mentioned in the previous sections, the end of inflation given by the
condition in Eq.~(\ref{endinflation}) does not occur on a uniform energy density
hypersurface~\cite{Naruko:2008sq},
i.e. the energy density is slightly different for the different end points.
For the $\delta N$ formalism to be valid,  $\delta N$ should be calculated at 
the uniform energy density hypersurface.
This is explained in \cite{Sasaki:2008uc} with the extra term $N_c$ in Eq.~(3.14)
of~\cite{Sasaki:2008uc}. They assume that the universe has become radiation-dominated right
after inflation. This means that at the time $t_e$ when 
the water fall field starts to become unstable, the energy density of the
inflaton fields changes to radiation instantly conserving the energy density, 
ignoring
 the details of the water fall field dynamics. If we define the time $t_c$
 sometime in the radiation dominated era, the change in the number of $e$--foldings between
$t_e$ and $t_c$ is
\dis{
N_c=\frac14 \ln \left[\frac{W_f}{W_0} \right].
}
This can be understood like this: During radiation domination,~\cite{Lyth:2003im}
\dis{
\frac{d\rho}{d N}=-4\rho,
}
which gives
\dis{
N_c=N(t_c)-N(t_e)=\frac14 \ln
\frac{\rho_e}{\rho_c}
=\frac14\ln\frac{W_f}{W_c}
=\frac14\ln \frac{W_f}{W_0}+\frac14\ln\frac{W_0}{W_c}.
}
where we can put $W_c=W_0$ which anyway does not change $\delta N$.
This is Eq.~(3.15) in~\cite{Sasaki:2008uc}. 
We require that the perturbation from this term,
$\delta N_c$, is smaller than from $\delta N_f$.
This was shown for the hybrid potential with a linear exponential dependence on the field in
\cite{Sasaki:2008uc}, but it was not shown in \cite{Naruko:2008sq} where they consider the same
hybrid potential as we do, with a quadratic dependence on the fields.
Therefore we need to check the validity of 
neglecting $\delta N_c$ in our two examples in the previous subsections.

In the first example with $g_2^2/g_1^2=\etap/\etac$, as in sec.~\ref{subsecA},
 the hypersurface at the end of inflation 
is a uniform energy density hypersurface, which means $\delta N_c=0$.
Thus in this case, there is no further change of $f_{NL}$,
if we assume an instantaneous change to radiation.

In the other case with $g_1^2=g_2^2$, as in sec.~\ref{subsecB}, we obtain
\dis{
N_c=\frac14\ln \frac{W_f}{W_0}= \frac18 \left(\etap \vp_{}^2+\etac \chi_{}^2 \right).
}
Taking the perturbation of $N_c$, we find 
\dis{
\delta N_c= \frac{\sigma^2}{8g_1^2}(\etac-\etap)\left[(\sin2\gamma)(\delta_1\gamma
+\delta_2\gamma)+(\cos2\gamma)(\delta_1\gamma)^2  \right],
}
where $\gamma$ is defined by $\tan\gamma=\chi/\vp$ at the end of 
inflation satisfying Eq.~(\ref{endinflation}).
Using the expressions for $\delta_1\gamma$ and $\delta_2\gamma$
((Eq (A2) and (A4) respectively in~\cite{Naruko:2008sq}), this becomes
\dis{
\delta N_c=&  \frac{\sigma^2}{8g_1^2}(\etac-\etap)
\left[\sin2\gamma\frac{-\etac\frac{\delta \vp_*}{\vp_*}+ \etap\frac{\delta
\chi_*}{\chi_*}}{\etac\tan\gamma+\etap/\tan\gamma}
+\frac{\sin2\gamma}{2}\frac{\etac\left(\frac{\delta \vp_*}{\vp_*}\right)^2
-\etap\left(\frac{\delta \chi_*}{\chi_*}\right)^2
}{\etac\tan\gamma+\etap/\tan\gamma}\right.\\
&+\left. \frac12\frac{\left\{\left(-\etac/\cos^2\gamma+\etap/\sin^2\gamma\right)\sin2\gamma +
\cos2\gamma(\etac\tan\gamma+\etap/\tan\gamma)\right\}\left(\etac\frac{\delta
\vp_*}{\vp_*}-\etap\frac{\delta
\chi_*}{\chi_*}\right)^2}{\left(\etac\tan\gamma+\etap/\tan\gamma\right)^3} 
 \right].\label{deltaNc}
}
The perturbation of $N$ from the flat hypersurface at horizon exit during 
inflation to the hypersurface at the end of the inflation 
(with a corrected factor of one half missing in
(A.8) in~\cite{Naruko:2008sq}),
\dis{
\delta N=  \frac{1/ \tan\gamma\frac{\delta \vp_*}{\vp_*}+ \tan\gamma\frac{\delta \chi_*}{\chi_*}
}{\etac\tan\gamma+\etap/\tan\gamma}
+\frac{1}{2}\frac{-1/\tan\gamma\left(\frac{\delta \vp_*}{\vp_*}\right)^2
+\tan\gamma\left(\frac{\delta
\chi_*}{\chi_*}\right)^2 }{\etac\tan\gamma+\etap/\tan\gamma}
+\frac{1/(\sin\gamma\cos\gamma)\left(\etac\frac{\delta \vp_*}{\vp_*}-\etap\frac{\delta
\chi_*}{\chi_*}\right)^2}{\left(\etac\tan\gamma+\etap/\tan\gamma\right)^3} .
\label{deltaN}
}

The final $\delta N_f$ from the flat hypersurface at horizon exit during 
inflation to the uniform energy density hypersurface during radiation dominated
era is the sum of both
\dis{
\delta N_f= \delta N +\delta N_c.
}

By comparing the corresponding term of $\delta\vp_*/\vp$, $\delta\chi_*/\chi$ etc between $\delta
N_c$ and $\delta N$,
we find that the term in $\delta N_c$ is suppressed compared to 
the corresponding term in  $\delta N$ by
\dis{
\frac{\sigma^2}{8g_1^2}(\etac-\etap) \times (\etap \quad \textrm{or} \quad \etac)
\times (\sin^2\gamma \quad \textrm{or} \quad \cos^2\gamma).
}
Here $\frac{\sigma^2}{g_1^2}=\varphi^2+\chi^2$.
Since we are working in the vacuum dominated regime and slow-roll, Eq.~(\ref{hybridsr}),
$\frac{\sigma^2}{g_1^2}\etap^2 \ll 1$ and $\frac{\sigma^2}{g_1^2}\etac^2 \ll 1$.
We therefore see that $\delta N_c$ is greatly suppressed compared to $\delta N$.

\subsection{Further evolution after inflation}

So far in this section we have assumed a quick transition
to the radiation epoch at the end of inflation, thereby neglecting the dynamics of the
waterfall field. However if we consider the role of the waterfall field, then
after the waterfall field is destabilised there may be a further evolution 
of the primordial curvature perturbation, 
which will lead to a change of the observable parameters. This applies
to any model with an inhomogeneous end of inflation since there are
 isocurvature perturbations still present after the waterfall field is 
destabilised and inflation has ended. Further evolution
will depend on the details of reheating in a model dependent way.
 To the best of our knowledge this 
issue has not been considered in depth in any paper on an inhomogeneous end of
 inflation. If we assume an instantaneous transition to radiation domination 
(so a completely efficient and immediate decay of the waterfall and 
inflaton fields) then there will be no further change to the
observables as we have argued in the previous section. 
However this is clearly an idealised case. For a review of reheating after inflation see for
example \cite{Bassett:2005xm}.

In the special case where the waterfall field is also light during inflation Barnaby and Cline
\cite{cline} have shown there is the possibility of generating a large non-Gaussianity during
preheating for certain parameter values. This is possible even if there is only one inflaton field
and the waterfall field present. However in this case inflation does not end abruptly when the
waterfall field is destabilised so this is not the scenario we have considered in this paper.

In practice the efficiency of reheating will depend on the couplings between the waterfall and
inflaton fields to any preheat fields, as well as on the ratio between $g_1,g_2$ and $\lambda$
\cite{GarciaBellido:1997wm}. In one regime where $\lambda\gg g_1$ and $g_2$ preheating depends mainly
on the inflaton fields. In the case where the two inflaton fields couple identically to all further
particles preheating depends on the coupling constants $g_1$ and $g_2$ and only much more weakly on
their bare masses proportional to $\etap$ and $\etac$ \cite{GarciaBellido:1997wm}. So in the case
where $g_1=g_2$ which we have considered earlier it may be reasonable to expect little or no further
evolution of the curvature perturbation, because the isocurvature perturbations should be irrelevant
and decay during preheating. On the other hand if $g_1/g_2\ll1$ which is the case considered in some
previous works on an inhomogeneous end to inflation the effect from preheating is more likely to be
important. In another regime where $\lambda\ll g_1$ and $g_2$ preheating depends mainly on the
waterfall field, so it may be that in this case the isocurvature perturbations are again
unimportant. However we should also consider the period from when the waterfall field is
destabilised and until preheating begins. It could be that the amount of expansion during this
time also leads to an extra correction to the curvature perturbation. This is a complicated issue
which deserves further investigation but is beyond the scope of this paper. To the best of our
knowledge there has been no study of preheating or reheating in a model of hybrid inflation with
more than one inflaton field.

\section{Scale dependence}
\label{sec:scale}

In general the bispectrum parameterised by $\fnl$ can have both a scale and a shape dependence. We
are considering the local form of $\fnl$ which means that $\fnl$ is shape independent. However it
can still have a (slow-roll suppressed) scale dependence \cite{Chen:2005fe,LoVerde:2007ri,Rath:2007ti}.

In our examples $\fnl$ has a scale dependence both because of the exponential term in $\fnl$,
(\ref{fNL_our}), and
because $\tr$ will vary through the change of the initial value of $\sis$. We find 
\bea \frac{\partial \ln \tr}{\partial \ln k}=\frac{\partial \ln e^{2 N(\etap-\etac)}}{\partial \ln
k} =-2(\etap-\etac). \eea

Using this we find from (\ref{fNL_our}) that
\bea n_{\fnl}-1 \equiv \frac{d\log \fnl}{d\log k}=
-4\frac{\etap-\etac}{1+\tr}. \eea 

In the case that we include the effect from the surface where the waterfall field is destabilised
and $g_1^2=g_2^2$ we find from (\ref{fNL_M}) that
\bea n_{\fnl}-1= -4\frac{\etap-\etac}{1+\left(\frac{\etap}{\etac}\right)^2\tr}. \eea

For both cases the spectral index of $\fnl$ satisfies 
\bea -4(\etap-\etac)<n_{\fnl}-1<0, \eea
 for any value of $\tr$ and hence $\fnl$
will be smaller on small scales
as we can see in the figures~\ref{hybridA} and~\ref{hybridB}.

Because we require a relatively large value of $\etap-\etac>1/N$ for our model to generate a large
non-Gaussianity it is quite possible for our model to generate a relatively significant scale
dependence of $\fnl$. However the amount also depends on $\tr$ and when this is large then
$n_{\fnl}-1$ is suppressed.

We note that this is in contrast to the large non-Gaussianity from an inhomogeneous end of
inflation found in~\cite{Naruko:2008sq}. In the specific cases they considered to generate a large
non-Gaussianity the non-Gaussianity was generated purely at the end of inflation and $\fnl$ is
scale independent.
In detail we see from Eqs.~(4.4) and~(4.24) in~\cite{Naruko:2008sq} that 
their formulae for $\fnl$ does not depend on $N$ or on any quantities 
evaluated at Hubble exit.

\section{Conclusion}
\label{sec:conclusion}

We have made an in depth study of a model of hybrid inflation with two 
inflaton fields (two-brid
inflation). We have studied the parameter space where non-Gaussianity may 
be large during
slow-roll inflation. In particular we have calculated the observable 
parameters $\Pz$,
$n_{\zeta}-1$, $r$, $\fnl$, $g_{NL}$ and $\tau_{NL}$. The spectral index
depends on the weighted sum of the slow-roll parameters $\etap$ and $\etac$ while the 
bispectrum depends
exponentially on the difference of the same slow-roll parameters, so it is 
possible to have a very
 large bispectrum and a scale invariant spectrum. We have shown that the 
trispectrum is generally
large through $\tau_{NL}$ whenever the bispectrum is large, but that the other 
parameter which
parameterises the trispectrum, $g_{NL}$ is always smaller than $\fnl$ and 
strongly suppressed
compared to $\tau_{NL}$. We have also shown that for certain initial conditions it is possible
that $\tau_{NL}$ is the only large non-linearity parameter, so the first observational signature of
non-Gaussianity from this model could be the trispectrum through $\tau_{NL}$.

Furthermore we have shown that during slow-roll inflation the loop corrections 
to the power
spectrum, $\fnl$ and $\tau_{NL}$ are always suppressed compared to the tree 
level terms. The
suppression follows from the constraint on the background trajectory that the 
classical background
value of the field should dominate over its quantum fluctuations.

We then investigated how the large non-Gaussianity which was generated during 
slow-roll may be
changed by the effects from the end of inflation. We included the waterfall field 
which ends inflation
when its effective mass becomes negative and it is destabilised. The effect of 
the waterfall field
depends on the values of the coupling constants between the waterfall field 
and the two inflaton
fields. If we choose the coupling constants so that surface where the 
waterfall field is
destabilised corresponds to a hypersurface of uniform energy density then 
there is no change to the
observables we calculated during slow roll. However if we choose the two 
coupling constants to be
equal then observables at both linear and higher order are changed by an amount 
that depends on the ratio $\etap/\etac$. 

\acknowledgements

The authors thank Yeinzon Rodriguez and Cesar Valenzuela-Toledo for numerous discussion on this
work. CB thanks Martin Bucher, Paolo Creminelli, Daniel Figueroa, Sarah Shandera and is especially
grateful
to Bartjan van Tent, Misao Sasaki and Filippo Vernizzi for interesting discussions about this work. 
CB and K.-Y.C acknowledge STFC and the University of Sheffield for
hospitality during a visit where part of this work was written.
CB acknowledges financial support from the Deutsche Forschungsgemeinschaft. K.-Y.C. is supported by
the Ministerio de Educacion y Ciencia of Spain under Proyecto Nacional FPA2006-05423 and
by the Comunidad de Madrid under Proyecto HEPHACOS, Ayudas de I+D S-0505/ESP-0346.  LMHH
acknowledges support from STFC.

\section{Appendix}

\subsection{Loop Suppression}
We require that the quantum fluctuation does not overwhelm the classical evolution of the fields.
In order to do that
we require  $|\dot{\chi}|\pi/H^2>\sqrt{3/2}$  for both fields on the
background trajectory~\cite{Creminelli:2008es}. 
With this constraint we show that the tree level dominates the loop correction.
In the slow-roll condition, this constraint leads to
\dis{
\left|\vp_*\right| > \sqrt{\frac{3}{2\pi^2}}\left|\frac{H_*}{\etap}\right|\qquad 
\textrm{and}\quad 
\left|\chi_*\right|   > \sqrt{ \frac{3}{2\pi^2}}\left|\frac{H_*}{\etac}\right|,
}
or equivelantly
\dis{
\left|\frac{\delta\vp_*}{\vp_*}\right| <\left| \frac{\etap}{\sqrt{6}}\right|,\qquad 
\left|\frac{\delta\chi_*}{\chi_*}\right| <\left| \frac{\etac}{\sqrt{6}}\right|\, . 
\label{classical}
}

Using the $\delta N$ formalism, 
\dis{
\delta N= N_\vp\delta\vp_* +N_\chi\delta\chi_* +\frac12N_{\vp\vp}(\delta\vp_*)^2
+\frac12N_{\chi\chi}(\delta\chi_*)^2 +N_{\vp\chi}(\delta\vp_*)(\delta\chi_*)
+\cdots,
}
in hybrid inflation, we find the first derivatives of the number of e-foldings
\dis{
N_\vp=\frac{\etap \vp_* e^{-2N\etap}}{2\epsilon},\qquad
N_\chi=\frac{\etac \chi_* e^{-2N\etac}}{2\epsilon},\label{firstd}
}
and the second derivatives
\dis{
N_{\vp\vp}&= \frac{N_\vp}{\vp_*}+\left(-4\etap+2\frac{\gamma}{\beta}\right)N_\vp^2,\\
N_{\chi\chi}&= \frac{N_\chi}{\chi_*}+\left(-4\etac+2\frac{\gamma}{\beta}\right)N_\chi^2,\\
N_{\vp\chi}&= 2 N_\vp N_\chi\left(\frac{\gamma}{\beta}-(\etap+\etac) \right),\label{secondd}
}
where
\dis{
\frac{\gamma}{\beta}= \frac{\etap^3 \vp_*^2 e^{-2N\etap}+ \etac^3 \chi_*^2
e^{-2N\etac}}{\etap^2 \vp_*^2 e^{-2N\etap}+ \etac^2 \chi_*^2 e^{-2N\etac} }.
}
We note that
\dis{
\left| \frac{\gamma}{\beta} \right|< \eta_{max}\equiv \textrm{max}\{\etap,\etac\}.
}

From the definition of the power spectrum and observations
\dis{
\calP_\zeta= \left(N_\vp^2+N_\chi^2 \right)\left(\frac{H_*}{2\pi}\right)^2 \simeq 10^{-10},
}
we know that
\dis{
\left|N_\vp \delta \vp\right| \simeq |N_\vp H_*| \lesssim 10^{-4},\qquad
\left|N_\chi \delta \chi\right| \simeq |N_\chi H_*| \lesssim 10^{-4}. \label{power}
}

Using Eq.~(\ref{classical}) and~(\ref{power}) 
we can easily check that the second order term in the expansion of $\delta N$ are always
smaller than the first order term. For example
\dis{
\frac12 N_{\vp\vp}(\delta\vp_*)^2= \frac12\frac{N_\vp}{\vp_*}(\delta\vp_*)^2
+\left(-2\eta_\vp+\frac{\gamma}{\beta}\right)(N_\vp\delta\vp_*)^2
<  \frac12 \left(\frac{\etap}{2\pi}\right)(N_\vp\delta\vp_*) +
 \eta_{max} 10^{-5} (N_\vp\delta\vp_*)<|\etap\Np|\delta\vp_*, \label{npp}
}
which is similar for the $N_{\chi\chi}$ term and 
\dis{
N_{\vp\chi}(\delta\vp_*\delta\chi_*)&= 2 (N_\vp\delta\vp_*)( N_\chi\delta
\chi_*)\left(\frac{\gamma}{\beta}-(\etap+\etac)
\right)<10^{-4}|\eta_{max}|\mathrm{max}(|N_{\vp}|\delta\vp_*,|N_{\chi}|\delta\chi_*).
}

The possible one loop domination can happen when $\chi$ is much smaller than $\delta\chi_*$, 
and $(\etap - \etac)N>1$ which is the case of Yeinzon et al.~\cite{Cogollo:2008bi}, but this can be possible only when we 
break the classical constraint we supposed in Eq.~(\ref{classical}).

In \cite{Cogollo:2008bi,Rodriguez:2008hy} the dominant loop correction to the power spectrum,
bispectrum and trispectrum was found to come from a single loop diagram in each case which is the
loop expansion one finds if truncating the $\delta N$ expansion at second order. Specifically the
loop corrections they found which may be dominant are (see figure~\ref{fig:234loop})
\bea
\Prat&=&\frac{N_{AB}N_{AB}}{(N_C N_C)^2}\Pz,  \\
\fNLl&=&\frac{5}{6}\frac{N_{AB}N_{BC}N_{AC}}{(N_D N_D)^3}\Pz, \\  
\tau_{NL}^{1\,\mathrm{loop}}&=& \frac{N_{AB}N_{BC}N_{CD}N_{AD}}{\left(N_E N_E\right)^4}\Pz.
\label{tauNL}
\eea
Similar formula for $\fnl$ and $\tau_{NL}$ at tree level are given in (\ref{NLformula})
(figure~\ref{fig:234tree}). By
comparing the terms at tree and 1--loop level (which are suppressed by a factor of
$\Pz\sim10^{-10}$) it is clear that the loop terms can only be large in the case that one of the
second derivative terms in the $\delta N$ expansion is extremely large. The integral over the
loop momentum gives rise to a logarythmic infrared divergence $\ln(kL)$ where $L$ is the large
scale cut off. We have followed \cite{Boubekeur:2005fj,Cogollo:2008bi,Rodriguez:2008hy} in assuming
that we can take $\ln(kL)\sim1$. In the terminology of \cite{Lyth:2007jh} this corresponds to
working in a minimal box. For more discussion of this divergence see for example 
\cite{Byrnes:2007tm,Seery:2007wf,Bartolo:2007ti,Enqvist:2008kt}.

Without loss of generality we assume $|\Npp|<\rm{max}(|\Npc|,|\Ncc|)$, so 
we need at least one of $|\Npc|/\Np^2\gg1$ or $|\Ncc|/\Np^2\gg1$ in order to have a
significant loop correction. First we see that the cross derivative term is never large, using
Eq.~(\ref{firstd}) and~(\ref{secondd}),
\bea 
\frac{|\Npc|}{\Np^2+\Nc^2}< 4\frac{\sqrt{\tr}}{1+\tr} |\eta_{max}| < 2|\eta_{max}|.
\eea

The ratio which can be large is $\Ncc/\Np^2$, in the case that it is large we have the simplified
formulas
\bea\label{Ploop} \Prat&\simeq&\left(\frac{\Ncc}{\Np^2}\right)^2\frac{1}{\left(1+\tr\right)^2}\Pz,
\eea
\bea
\fNLt&\simeq&\frac56 \frac{\Ncc}{\Np^2}\frac{\tr}{\left(1+\tr\right)^2}, \qquad \fNLl\simeq \frac56
\left(\frac{\Ncc}{\Np^2}\right)^3\frac{1}{\left(1+\tr\right)^3}\Pz, \label{fNLloop}
\eea
\bea 
\tau_{NL}^{\mathrm{tree}}&\simeq& \left(\frac{\Ncc}{\Np^2}\right)^2\frac{\tr}{\left(1+\tr\right)^3},
\qquad \tau_{NL}^{1\,\mathrm{loop}}\simeq
\left(\frac{\Ncc}{\Np^2}\right)^4\frac{1}{\left(1+\tr\right)^ 4}\Pz. \label{tauNLloop}\eea
We recall the definition $\tr\equiv(N_{\chi}/N_{\vp})^2$. So it appears that by making $\Ncc/\Np^2$
large enough we can make the loop corrections as large as we like.

However using Eq.~(\ref{power}) and~(\ref{npp}), we find the inequality
\bea
\left(\frac{\Ncc}{\Np^2}\right)^2\frac{\Pz}{\tr(1+\tr)}<\etac^2\ll1,
\eea
and hence we see from (\ref{Ploop}) (\ref{fNLloop}) and (\ref{tauNLloop}) that the loop correction
to the power
spectrum, $\fnl$ and $\tau_{NL}$ is always suppressed.

\subsection{Other loop corrections}

So far we have proved that the loop corrections which \cite{Cogollo:2008bi,Rodriguez:2008hy} found
to be largest are still suppressed compared to the tree level terms. Here we show that these
particular loop corrections remain the dominant loop corrections even for a more general trajectory
in field space. To do this we need to go beyond second order in the $\delta N$ expansion. In fact
to prove that all of the 1--loop corrections to the power spectrum, $\fnl$ and $\tau_{NL}$ are
suppressed we need to go up to fourth order in the $\delta N$ expansion.

At third order, the only derivatives which can be very large are
\bea\label{N3rd} \frac{\Npcc}{\Np^3}=-2\etac \frac{\Ncc}{\Np^2}, \qquad
\frac{\Nccc}{\Np^3}=6\sqrt{\tr}(\etap-2\etac)\frac{\Ncc}{\Np^2}, \eea
while at fourth order the largest derivative is given by
\bea\label{N4th} \frac{\Ncccc}{\Np^4}=6(\etap-2\etac)\left(\frac{\Ncc}{\Np^2}\right)^2. \eea
We use the third order derivatives to calculate $g_{NL}$, see (\ref{NLformula}) and
(\ref{NLtrispectrum}).

\begin{figure}
\scalebox{1.3}{\includegraphics*{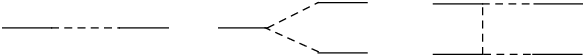}} \caption{Tree level diagrams for
the power spectrum (left hand side), the bispectrum (centre) and the relevant tree diagram for the
trispectrum (right) which corresponds to $\tau_{NL}$. The diagrams were drawn
using JAXODRAW \cite{Binosi:2003yf}}\label{fig:234tree}
\end{figure}

\begin{figure}
\scalebox{1.3}{\includegraphics*{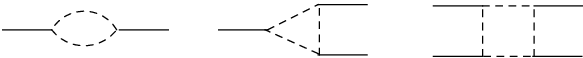}} \caption{The dominant one-loop level diagrams for
the power spectrum (left hand side), the bispectrum (centre) and the trispectrum
(right).}\label{fig:234loop}
\end{figure}

For details of the loop corrections we refer the reader to \cite{Byrnes:2007tm}, in which a
diaggrammatic approach to calculating the primordial $n$--point functions of $\zeta$ at any required
loop level was developed. The formula for the $n$--point functions depend on the derivatives in the
$\delta N$ formalism, for every vertex with $r$ internal legs there is a corresponding term with $r$
derivatives of $N$. We note from Fig.~\ref{fig:234tree} that the tree level diagrams for the power
spectrum, $\fnl$ and
$\tau_{NL}$ all depend on diagrams with vertices connected to either one or two internal lines. 
The dominant loop correction found by
\cite{Cogollo:2008bi,Rodriguez:2008hy} in each case follows by adding an internal line between the
two vertices with a single internal line attached in the tree level diagrams. These diagrams are
 shown in Fig.~\ref{fig:234loop}. This converts two
first order derivatives of $N$ into second order derivatives, which adds a large multiplicative
factor of $(\Ncc/\Np^2)^2$ to the 1--loop term. When constructing any other 1--loop diagram we can
either dress a vertex, which means adding two internal lines to a single vertex or connect an
internal line between two external vertices where at least one of them does not have a single
internal line atteched at tree level. In each of these cases it follows from (\ref{N3rd}) and
(\ref{N4th}) that these other diagrams are not boosted by a factor larger than $\Ncc/\Np^2$ and
hence are subdominant to the loops found in \cite{Cogollo:2008bi,Rodriguez:2008hy}.

We conclude that all one-loop terms to the power spectrum, $\fnl$ and $\tau_{NL}$ are suppressed
compared to the tree level terms, even for a general trajectory in field space. In fact this
argument applies to any loop term, provided that we can truncate the expansion of $\delta N$ at
fourth order. When drawing an $l+1$--loop level diagram we add an extra internal line to a diagram
at $l$--loop level. This means attaching in total two extra derivatives to the derivatives of $N$
corresponding to the vertices where the extra internal line is added. At most this can add a
numerical factor of $(\Ncc/\Np^2)^2$. However there is also a suppression factor of order $\Pz$
which comes from going to a higher order in loops, which as we have seen explicitly in the case of
the one loops diagrams always leads to a suppression.



\end{document}